Comment on

"Examining the effect of counternarratives about physics on women's physics career intentions"

M. B. Weissman[1] and J. M. Robins[2]

[1]Department of Physics, University of Illinois at Urbana-Champaign

1110 West Green Street, Urbana, IL 61801-3080

[2]Departments of  Epidemiology and  Biostatistics
Harvard T.H. Chan School of Public Health
677 Huntington Av., MA Boston 02115

Abstract: A paper evaluating the effects of lessons intended to encourage high school students to continue physics studies made some important errors. One was to underestimate the width of confidence intervals by failing to use standard cluster randomization analysis. Another was to use a missing-data imputation program that inappropriately assumes that data are missing at random, leading to potential bias in estimating the effect. The last was to omit discussion of how the treatment used was likely to produce substantial social desirability survey response bias, eroding external validity.

Physics education research can help to achieve educational goals by providing estimates of the effects of various changes in educational practices. A recent paper in PRPER commendably compared effects of different actual interventions(*1*), a more promising approach than attempts to infer such causal effects from correlations that are insufficient to identify them. (*2, 3*) The goal was to see if some specific lessons increased high school students' propensity to pursue further physics education. (*1*) Unfortunately, the paper had some major errors in statistical analysis and did not discuss an important caveat concerning its external validity. This Comment is intended to point out these problems, some of which could be repaired by re-analysis and others ameliorated in future research.

The study assigned lessons on careers in physics and women in physics to a treatment group of classes. (*1*) Classes in a control group were assigned more traditional lessons on major accomplishments in physics *per se*. The assignment was described as "random", although the randomization process was not specified. Students were surveyed on their intent for future physics study and careers both before and after the lessons. After an initial exploration without a control group, ten new teachers were picked for the treatment group and three were picked for the control group. Changes in survey response were compared between the treatment and control groups, including analyses on subgroups such as females and under-represented minorities. To fill in the 38% of the data that were missing due to "the challenges of collecting data multiple times during the semester from students in real high school classrooms" (*1*), the analysis used a multiple imputation program that assumes that those data are missing at random.(*4*)

The paper compared treatment and control changes for four groups- female, non-female, minority, and non-minority. (*1*) The differences between the treatment and control groups are reported in Table V as Cohen d's of 0.23, 0.13, 0.17, and 0.20, with 95% confidence intervals (0.07,0.39), (-0.03,0.30), (0.05, 0.30) and (-0.05, 0.44), respectively.

The calculation of the confidence intervals and p-values failed to use the cluster-randomization methods required for any such experiment in which the differences between treatment groups (schools, classes, and especially teachers) can contribute substantially to the between-group variance, beyond the variance obtained from individual student-level variation.(*5*) The importance of including the cluster term is routinely recognized, e.g. in medical literature.(*6*) This extra variance would make the confidence intervals wider and the p-values larger than those given in the paper.

The need for considering cluster effects can easily be seen in a limiting case. Say that the treatment group consisted of one teacher's classes at one school and the control group consisted of another teacher's classes at another school. Even if within-group variance were small, observed differences in outcomes between the two groups could be due to the different types of students at the two schools or to the different teachers. Thus in this extreme case treatment effects could not be identified. With multiple teachers identification is possible but with enhanced uncertainty. With only three teachers in the control group, the actual study(*1*) was not far from the limiting case.

How important is the cluster effect in this case? Even aside from differences between teachers themselves, Table XI shows differences between the treatment and control groups well beyond those expected from individual-level differences. (*1*) For example, the difference between the female treatment and control groups before the treatment starts is more than four times the standard error expected from individual differences, a difference substantially larger than the subsequent treatment effect difference between the groups. Neither the sign nor magnitude of the effects of such large initial differences between the groups on the subsequent changes is known.

If the effect sizes were given for each of the ten teachers in the treatment group of the controlled study one could use inter-teacher variance to estimate the net variance, since that would include both the student-level and teacher-level terms. One could then provisionally and generously assume that the same variance applied in the three-teacher control group and apply standard multi-level modeling(*7*). (Trying to separately infer inter-teacher variance in that control group with two degrees of freedom would almost certainly lead to loss of nominal significance.) That such estimates must be made from a relatively small number of teachers also increases the confidence intervals and the p-values because the number of degrees of freedom in the corresponding t-tests used would be greatly reduced. Unfortunately, the effect sizes for individual teachers were not given, so readers cannot do the calculation of the actual cluster confidence intervals.(*1*) With such small numbers of independent data points and with an unknown distribution of teacher effects a non-parametric rank test on the mean effects in the 13 groups with different teachers might be more appropriate than a t-test, which assumes normality of the underlying distribution.

The large fraction of missing data creates another problem, perhaps more severe because it directly affects the point estimates, not just the error bars. If the students who were unconvinced by the lessons were particularly likely not to fill out later surveys the effect estimate for complete cases would be biased upward. The software used to fill in the missing data is not designed to fix this bias. The software description cited says "Amelia assumes, as most multiple imputation methods do, that the data are missing at random (MAR). This assumption means that the pattern of missingness only depends on the observed data $D^{obs}$, not the unobserved data $D^{mis}$. "(*4*) In this case that MAR imputation assumption seems unrealistic, since failure to respond on the second survey is likely to be closely related to the effects of the treatment in ways that cannot be fully predicted using the few observed variables.

With 38% of the data points missing, a majority of the pairwise start-to-finish changes are likely to have been unmeasured since one needs *both* start and finish values to know the difference. Thus biased estimation of the unmeasured changes due to the inappropriate MAR assumption might make a substantial difference in the results. Although the paper says "the multiply imputed inferences are much more robust" than the "raw, unimputed findings" the robustness may be an artifact of the MAR assumption. In simple terms, the imputation becomes more robust by adding data that are assumed to look like the observed data, but that assumption is unrealistic here. One would need a fuller description of which data were missing and of the parameters of the imputation procedure to make a rough estimate of how much such effects might bias the point estimates and confidence intervals.

Regardless of whether nominal statistical significance would survive in a standard treatment of the statistics, the external validity of the point estimates of the treatment effect is questionable. The authors acknowledge that one cannot confidently extrapolate from survey response changes in high school to behavior changes downstream.(*1*) The problem may be more severe than that generic caveat suggests. The paper does not indicate that the responses were anonymous and the results are at points analyzed in terms of paired t-tests. (*1*) That indicates that individual student responses could be tracked. Lack of anonymity might exacerbate the general problem of social-desirability bias(*8*), i.e. students, especially females(*9*), will tend to give answers that teachers desire regardless of their actual long-term intent.

How strong might the social desirability effects be in this particular study? The paper(*1*) gives a link (*10*) to the materials used for the lessons, including a video of a teacher giving a Careers in Physics lesson. At approximately 5 minutes into the video, the teacher points to an "I [heart] Physics" sign and tells the class:

> "Remember, show your love for physics. Some of you did. And you're going to make me happy."

It is difficult to imagine that prudent students would fail to adjust their answers toward the teacher's desires when a follow-up survey is taken in the same teacher's class. Thus, regardless of the statistical issues, it is unclear how much if any of the modest survey effect reported would indicate a real effect on long-term intentions.

Some of the issues raised here might easily be cleared up using existing data. The cluster confidence intervals could be calculated from the results broken-out by teacher. The results of a complete-case analysis along with some characterization of which cases were most likely to be incomplete and more detailed description of the imputations might help to get an informal sense of how important non-response bias might be. It is difficult to see how one might retroactively correct for possibly intense social-desirability bias, but in future work it could perhaps be ameliorated by having a set script for both treatment and control teachers to ask students to respond realistically on the surveys, along with explicit instructions to teachers not to plead with the students to "show love for physics" on their responses. For each issue, having a well-specified pre-registered protocol would be useful.(*11*)

For the project described in this paper getting methods right has more immediate practical importance than for typical academic papers. A large-scale project for implementing high school classroom changes based on this study is already being rolled out. (*10*) Such projects are not cost free. Somewhere between 120 and 180 minutes of class time is required for the lessons, with additional time required for homework assignments and significant investment of teacher time.(*10*) The materials presented to the students include a claim that the high MCAT and LSAT scores obtained by physics students show that physics helps them "gain skills that give them a competitive edge for Medical and Law School", (*10*) a poor example of reasoning about causation. Students going into a wide range of fields might benefit most from lessons on careful reasoning about data and from researchers setting good examples of such reasoning.